\documentclass[preprint,showpacs,preprintnumbers]{revtex4}
\usepackage{amsfonts}
\usepackage{amsmath}
\usepackage{graphicx}
\usepackage{dcolumn}
\usepackage{bm}
\usepackage{epsfig}

\begin{document}

\title{\textbf{COMPRESSIBILITY EFFECTS ON THE LIGHT SCATTERED BY A
NON-EQUILIBRIUM SUSPENSION IN A NEMATIC SOLVENT}}
\author{H. H\'{\i}jar}
\author{R. F. Rodr\'{\i}guez}
\email{zepeda@fisica.unam.mx}
\altaffiliation[Also at ]{FENOMEC. Fellow of SNI, Mexico.}
\affiliation{Instituto de F\'{\i}sica. Universidad Nacional Aut\'{o}noma de M\'{e}xico.\\
Apdo. Postal 20-364, 01000 M\'{e}xico, D. F., M\'{e}xico.}
\date{}

\begin{abstract}
We investigate the effects produced on the light scattering spectrum by the
anisotropic diffusion of impurities (dye) in a compressible nematic solvent.
This spectrum is calculated by using a fluctuating hydrodynamic description
when the system is in both, a fully thermodynamic equilibrium state and in a
non-equilibrium steady state ($NESS$) induced by a dye-concentration
gradient. In the former state, the isotropic pre-transitional as well as the
nematic phase of the solvent are considered. We find that the equilibrium
spectrum is symmetric (Lorentzian) with respect to the frequency shifts, but
anisotropic through its explicit dependence on the ratio of the parallel and
normal diffusion coefficients of the dye. The values of these coefficients
were taken from experimental measurements of diffusion of methylred and
nitrozo di-methyl aniline in a $MBBA$ solvent. We find that the
compressibility of the solvent increases the maximum and the width at half
height of the Rayleigh peak, with respect to the incompressible case \cite%
{suspension1}. This increase varies between $12\%$ and $25\%$, respectively,
when the impurities concentrations is the range of $1\%$ - $5\%$. The $NESS$
induces a coupling between the concentration fluctuations of the dye and the
hydrodynamic fluctuations of the solvent. In this case the compressibility
effects may increase the maximum and the width of the central peak up to $%
25\%$, for values of the concentration gradient four orders of magnitude
smaller than those considered in the incompressible case. This result
indicates that compressibility and mode coupling effects may be significant
and that they might be detected experimentally. On the other hand, for the
nonequilibrium Brillouin part of the spectrum we find that the intensities
of the sound propagation modes are unequal and one of the peaks shrinks in
the same amount as the other increases. This asymmetry increases linearly
with the magnitude of the solute concentration gradient. The maximum
difference between the nonequilibrium and equilibrium contributions to the
Brillouin spectrum for various values of the external gradient is also
estimated. However, in all cases we find that the amplitude of the Brillouin
peaks is several orders of magnitude smaller than the central peak.
Therefore, although nonequilibrium effects do produce an asymmetry, our
theoretical analysis indicates that these effects are too small to be
observed experimentally.
\end{abstract}

\maketitle

\section{Introduction}

In a previous paper hereafter referred to as $I$ \cite{suspension1}, we have
analyzed the effects produced by the anisotropic diffusion of impurities in
an incompressible nematic solvent. By using a fluctuating hydrodynamic
approach we calculated the central Rayleigh peak of the spectrum when the
solvent is in both, a fully thermodynamic equilibrium state and in a
non-equilibrium steady state $(NESS)$ induced by a dye-concentration
gradient.\ The nonequilibrium states considered in $I$ were close to
equilibrium and the driving concentration gradient was taken into account
through a local version of the fluctuation-dissipation theorem for the
stochastic current of the impurities. In the present work we extend the
analysis of $I$ in two aspects. First, instead of introducing spatial
inhomogeneities in the fluctuation-dissipation theorem, we consider mode
coupling terms relating the concentration fluctuations of the solute and the
orientation and velocity fluctuations of the solvent. Secondly, apart from
the central peak, in the present work we also calculate the Brillouin part
of the dynamic structure factor of the impurities and analyze the mode
coupling contribution to the full spectrum of the suspension. For this
purpose the compressibility of the solvent is considered explicitly in its
equations of motion.

We find that in equilibrium the compressibility of the solvent increases the
maximum and the width at half height of the Rayleigh peak, with respect to
the incompressible case, by amounts that vary between $12\%$ and $25\%$,
respectively, as a function of the impurities concentrations in the range of
$1\%$ - $5\%$. In the $NESS$ induced by the concentration gradient, these
features of the central peak may increase up to $25\%$, even for values of
the concentration gradient which are four orders of magnitude smaller than
those considered in $I$. This result indicates that compressibility and mode
coupling effects may be significant and that they might be detectable
experimentally.

On the other hand, for the Brillouin part of the spectrum in the $NESS$ we
find that the intensities of the sound propagation modes are unequal and one
of the peaks shrinks in the same amount as the other increases, a behavior
that is also predicted and observed for a simple fluid \cite{beysens}, \cite%
{sengers}, \cite{law1}. We find that this asymmetry increases linearly with
the magnitude of the solute concentration gradient. The maximum difference
between the nonequilibrium and equilibrium contributions to the Brillouin
spectrum for different values of the external gradient is also estimated.
However, in all cases we find that the amplitude of the Brillouin peaks is
several orders of magnitude smaller than that of the central peak.
Therefore, although nonequilibrium effects are present in the above
mentioned asymmetry, our theoretical analysis indicates that these effects
are too small and difficult to observe experimentally.

\section{Model and Basic Equations}

As in \emph{I}, we consider a dilute suspension of noninteracting impurities
diffusing through a thermotropic nematic liquid crystal solvent, as depicted
in Fig. 1. If the suspension is sufficiently diluted, the dynamics of the
impurities does not disturb appreciably the state of the nematic and it may
be considered to be in an equilibrium state defined by a temperature $T_{o}$%
, a pressure $p_{o}$, a vanishing velocity field
$\overrightarrow{v}^{o}=0$ and a uniform director's orientation
$\hat{n}^{o}=\left( 0,0,1\right) $, corresponding to the
homeotropic configuration shown in Fig. 1. X
\begin{figure}[tbp]
\begin{minipage}{\textwidth}
\includegraphics[ width=11cm]{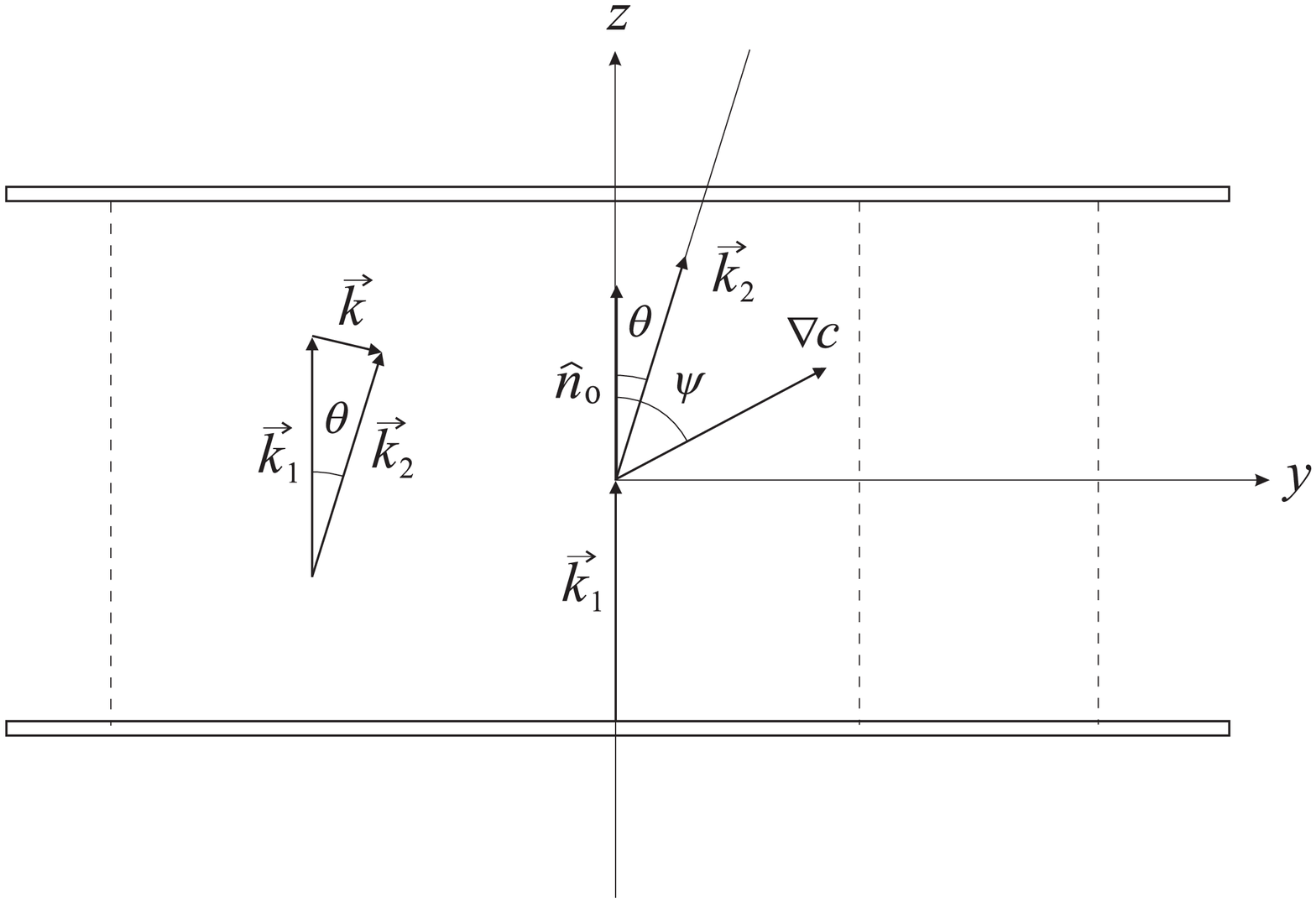}
\caption{Schematic representation of a plane homeotropic cell with a
constant thermal gradient along $z$ direction. The inset shows the
scattering geometry.The scattering angle is $\protect\theta $.}
\end{minipage}
\end{figure}

\bigskip We shall only consider nonequilibrium states corresponding to a
stationary concentration field of impurities defined by
\begin{equation}
c^{s}\left( \overrightarrow{r}\right) =c_{o}+\overrightarrow{r}\cdot
\overrightarrow{a},  \label{2.1}
\end{equation}%
where $c_{o}$ is the mean concentration of impurities and $\overrightarrow{a}%
\equiv \nabla c$ is the uniform concentration gradient in the $y-z$ plane
and whose direction is specified by $\psi $. For future use it will be
convenient to recast (\ref{2.1}) in the more convenient form
\begin{equation}
c^{s}\left( \overrightarrow{r}\right) =c_{o}+a\lim_{q\longrightarrow 0}\frac{%
\sin \left( \overrightarrow{q}\cdot \overrightarrow{r}\right) }{q},
\label{2.6}
\end{equation}%
where $a\equiv \left\vert \overrightarrow{a}\right\vert $, $\overrightarrow{q%
}$ $\equiv \frac{1}{a}\nabla c$ is an auxiliary vector of variable magnitude
and parallel to $\overrightarrow{a}$.

If no chemical reactions occur between the impurities, their total number is
conserved and their local concentration density, $c\left( \overrightarrow{r}%
,t\right) $, obeys the continuity equation
\begin{equation}
\frac{\partial c}{\partial t}+\nabla _{i}J_{i}=0,  \label{2.2}
\end{equation}%
where $J_{i}\left( \overrightarrow{r},t\right) $ is the flux of the
diffusing particles, which for an uniaxial nematic is of the form
\begin{equation}
J_{i}\left( \overrightarrow{r},t\right) =-D_{ij}\left( \overrightarrow{r}%
,t\right) \nabla _{j}c\left( \overrightarrow{r},t\right) +c\left(
\overrightarrow{r},t\right) v_{i}\left( \overrightarrow{r},t\right) .
\label{2.3}
\end{equation}%
The first term on the right hand side is the usual Fick's law contribution
where $D_{ij}\left( \overrightarrow{r},t\right) $ is the diffusion tensor of
the suspended impurities; the second term represents the convective
diffusion of the impurities, where $v_{i}\left( \overrightarrow{r},t\right) $
denotes the velocity field of the solvent. For an uniaxial nematic $D_{ij}$
has the standard form%
\begin{equation}
D_{ij}\left( \overrightarrow{r},t\right) =D_{\bot }\delta _{ij}+\left(
D_{\parallel }-D_{\bot }\right) n_{i}\left( \overrightarrow{r},t\right)
n_{j}\left( \overrightarrow{r},t\right) ,  \label{2.4}
\end{equation}%
where $D_{\parallel }$ is the diffusion coefficient parallel to the director
$\widehat{n}\left( \overrightarrow{r},t\right) $ and $D_{\perp }$ is the
corresponding coefficient in the perpendicular direction. $D_{a}\equiv
D_{\Vert }-D_{\bot }$ is the corresponding diffusion anisotropy. Usually,
the diffusion of small particles dissolved in a nematic solvent is such that
the diffusion parallel to the director is faster than perpendicular to it;
as a consequence, the ratio $D_{\Vert }/D_{\bot }$ seems to be independent
of the actual shape of the diffusing molecules \cite{rondelez}, \cite{janig}%
, \cite{franklin}. Using Eq. (\ref{2.4}), the diffusion equation for $c(%
\overrightarrow{r},t)$ turns out to be%
\begin{eqnarray}
\frac{\partial }{\partial t}c\left( \overrightarrow{r},t\right)  &=&D_{\bot
}\nabla ^{2}c+D_{a}n_{i}n_{j}\nabla _{i}\nabla _{j}c  \notag \\
&&+D_{a}\left( n_{i}\nabla _{i}n_{j}+n_{j}\nabla _{i}n_{i}\right) \nabla
_{j}c  \notag \\
&&-v_{i}\nabla _{i}c+c\nabla _{i}v_{i}.  \label{2.4a}
\end{eqnarray}%
Note that the compressibility of the solvent is taken into account by the
term $\nabla _{i}v_{i}\neq 0$, which was absent in $I$.

If a fluctuating mass diffusion current, $J_{i}^{F}\left( \overrightarrow{r}%
,t\right) $, is introduced into this equation, the concentration
fluctuations $\delta c(\overrightarrow{r},t)\equiv c(\overrightarrow{r}%
,t)-c^{s}\left( \overrightarrow{r}\right) $, obey the linearized equation
\begin{eqnarray}
\frac{\partial }{\partial t}\delta c &=&\left( D_{\bot }\nabla _{\bot
}^{2}+D_{\Vert }\nabla _{z}^{2}\right) \delta c+a_{i}\delta
v_{i}-c^{s}\nabla _{i}\delta v_{i}  \notag \\
&&+D_{a}\left( a_{i}\nabla _{z}\delta n_{i}+a_{z}\nabla _{i}\delta
n_{i}\right) -\nabla _{i}J_{i}^{F}.  \label{2.8}
\end{eqnarray}%
$J_{i}^{F}\left( \overrightarrow{r},t\right) $ is a Markovian, Gaussian,
stochastic processes with zero mean $\left\langle J_{i}^{F}\left(
\overrightarrow{r},t\right) \right\rangle =0$ and whose correlation is
assumed to obey a local equilibrium version of the usual
fluctuation-dissipation relation \cite{landau},
\begin{equation}
\left\langle J_{i}^{F}\left( \overrightarrow{r},t\right) J_{j}^{F}\left(
\overrightarrow{r}^{\prime },t^{\prime }\right) \right\rangle
=2D_{ij}c_{s}\left( \overrightarrow{r}\right) \delta \left( \overrightarrow{r%
}-\overrightarrow{r}^{\prime }\right) \delta \left( t-t^{\prime }\right) .
\label{2.9}
\end{equation}%
If we define the Fourier transform of an arbitrary field $f\left(
\overrightarrow{r},t\right) $ by
\begin{equation}
\widetilde{f}\left( \overrightarrow{k},\omega \right) \equiv \int \int
dtd^{3}r\text{ }f\left( \overrightarrow{r},t\right) e^{-i\left(
\overrightarrow{k}\cdot \overrightarrow{r}-\omega t\right) },  \label{2.10}
\end{equation}%
Eq. (\ref{2.8}) reads%
\begin{eqnarray}
\delta \tilde{c}\left( \overrightarrow{k},\omega \right)  &=&G\left(
\overrightarrow{k},\omega \right) \left[ -ic_{o}k_{i}\delta \tilde{v}_{i}+%
\overrightarrow{a}\cdot \nabla _{\overrightarrow{k}}\left( k_{i}\delta
\tilde{v}_{i}\right) -a_{i}\delta \tilde{v}_{i}\right.   \notag \\
&&\left. +iD_{a}\left( k_{z}a_{i}\delta \tilde{n}_{i}+a_{z}k_{i}\delta
\tilde{n}_{i}\right) -ik_{i}\tilde{J}_{i}^{F}\right] ,  \label{2.11}
\end{eqnarray}%
where the propagator $G\left( \overrightarrow{k},\omega \right) $ is given by%
\begin{equation}
G\left( \overrightarrow{k},\omega \right) =\left( -i\omega +D_{\perp
}k_{\perp }^{2}+D_{\parallel }k_{z}^{2}\right) ^{-1},  \label{2.12}
\end{equation}%
with $k_{\perp }^{2}\equiv k_{x}^{2}+k_{y}^{2}$ and where $\nabla _{%
\overrightarrow{k}}$ denotes the differential operator in $\overrightarrow{k}
$-space, $\nabla _{\overrightarrow{k}}\equiv \left( \partial /\partial
k_{x},\partial /\partial k_{y},\partial /\partial k_{z}\right) $. Similarly,
Eq. (\ref{2.9}) becomes%
\begin{eqnarray}
\left\langle \tilde{J}_{i}^{F}\left( \overrightarrow{k},\omega \right)
\tilde{J}_{j}^{F}\left( \overrightarrow{k}^{\prime },\omega ^{\prime
}\right) \right\rangle  &=&2\left( 2\pi \right) ^{4}D_{ij}\delta \left(
\omega +\omega ^{\prime }\right) \left\{ c_{o}\delta \left( \overrightarrow{k%
}+\overrightarrow{k}^{\prime }\right) \right.   \notag \\
&&\left. +\frac{a}{2i}\left[ \delta \left( \overrightarrow{k}+%
\overrightarrow{k}^{\prime }-\overrightarrow{q}\right) -\delta \left(
\overrightarrow{k}+\overrightarrow{k}^{\prime }+\overrightarrow{q}\right) %
\right] \right\} .  \label{2.12a}
\end{eqnarray}

\subsection{Impurities Structure Factor in Equilibrium}

The impurities dynamic structure factor in equilibrium, $S^{eq}\left(
\overrightarrow{k},\omega \right) $, is obtained by setting $\overrightarrow{%
a}=0$ in Eqs. (\ref{2.11}), (\ref{2.12a}) to yield%
\begin{eqnarray}
S^{eq}\left( \overrightarrow{k},\omega \right)  &\equiv &\left\langle \delta
\tilde{c}\left( \overrightarrow{k},\omega \right) \delta \tilde{c}\left( -%
\overrightarrow{k},-\omega \right) \right\rangle   \notag \\
&=&S_{J}^{eq}\left( \overrightarrow{k},\omega \right) +S_{vv}^{eq}\left(
\overrightarrow{k},\omega \right)   \notag \\
&=&2\left( 2\pi \right) ^{4}\left\vert G\left( \overrightarrow{k},\omega
\right) \right\vert ^{2}\delta ^{4}\left( 0\right) c_{o}k_{i}k_{j}D_{ij}
\notag \\
&&-\left\vert G\left( \overrightarrow{k},\omega \right) \right\vert
^{2}c_{o}^{2}k_{i}k_{j}\left\langle \delta \tilde{v}_{i}\left(
\overrightarrow{k},\omega \right) \delta \tilde{v}_{j}\left( -%
\overrightarrow{k},-\omega \right) \right\rangle .  \label{2.13}
\end{eqnarray}%
The first term is the contribution due to the stochastic current $J_{i}^{F}$
and the second one is the contribuiton arising from the dynamics of the
nematic solvent through its velocity correlation functions. Note that in
contrast to $I$, the equilibrium correlation function of the solvent
velocity fluctuations appears explicitly in this equilibrium property of the
solute. This correlation will now be calculated from the fluctuating
hydrodynamic equations for the solvent.

The hydrodynamic state of the nematic is described in terms of the pressure $%
p\left( \overrightarrow{r},t\right) $, temperature, $T(\overrightarrow{r},t)$
and the velocity $\overrightarrow{v}(\overrightarrow{r},t)$ fields, and the
unit vector defining the local symmetry axis (director), $\hat{n}(%
\overrightarrow{r},t)$. To calculate $\left\langle \delta \tilde{v}%
_{i}\left( \overrightarrow{k},\omega \right) \delta \tilde{v}_{j}\left( -%
\overrightarrow{k},-\omega \right) \right\rangle $ and other correlation
functions of the solvent that will be appear below, it is convenient to
separate the state variables into two independent sets, namely, transverse
and longitudinal variables with respect to the plane $\hat{n}^{o}-%
\overrightarrow{k}$, \cite{forster}, \cite{forster2}. The former set is $%
\left\{ \delta \tilde{n}_{1},\delta \tilde{v}_{1}\right\} $ with
\begin{equation}
\delta \tilde{n}_{1}\equiv k_{\perp }^{-1}\hat{n}^{o}\cdot \left(
\overrightarrow{k}\times \delta \widetilde{\overrightarrow{n}}\right) ,
\label{2.18b}
\end{equation}%
\begin{equation}
\delta \tilde{v}_{1}\equiv k_{\perp }^{-1}\hat{n}^{o}\cdot \left(
\overrightarrow{k}\times \delta \widetilde{\overrightarrow{v}}\right) ,
\label{2.19a}
\end{equation}%
while the longitudinal set is $\left\{ \delta \tilde{n}_{3},\delta \tilde{v}%
_{2},\delta \tilde{T},\delta \tilde{v}_{3},\delta \tilde{p}\right\} $ with%
\begin{equation}
\delta \tilde{v}_{2}\equiv k^{-1}k_{\perp }^{-1}\overrightarrow{k}\times %
\left[ \overrightarrow{k}\times \hat{n}^{o}\right] \cdot \delta \widetilde{%
\overrightarrow{v}},  \label{2.20}
\end{equation}%
\begin{equation}
\delta \tilde{v}_{3}\equiv k^{-1}\overrightarrow{k}\cdot \delta \widetilde{%
\overrightarrow{n}}  \label{2.21}
\end{equation}%
and
\begin{equation}
\delta \tilde{n}_{3}\equiv k^{-1}\overrightarrow{k}\cdot \delta \widetilde{%
\overrightarrow{n}},  \label{2.22}
\end{equation}%
Substitution of these definitions in Eq. (\ref{2.13}), evaluation of the
resulting expression at $\overrightarrow{k}^{\prime }=-\overrightarrow{k}$, $%
\omega ^{\prime }=-\omega $ and use of the explicit expression of the
propagator $G\left( \overrightarrow{k},\omega \right) $, leads to the
following equilibrium structure factor
\begin{equation}
S^{eq}\left( \overrightarrow{k},\omega \right) =2\left( 2\pi \right)
^{4}\delta ^{4}\left( 0\right) c_{o}\frac{\omega _{D}(\overrightarrow{k})}{%
\omega ^{2}+\omega _{D}^{2}(\overrightarrow{k})}\left\{ 1-\frac{c_{o}k^{2}}{%
\omega _{D}(\overrightarrow{k})}\left\langle \delta \tilde{v}_{3}\left(
\overrightarrow{k},\omega \right) \delta \tilde{v}_{3}\left( -%
\overrightarrow{k},-\omega \right) \right\rangle \right\}   \label{2.23}
\end{equation}%
with%
\begin{equation}
\omega _{D}\left( \overrightarrow{k}\right) \equiv D_{\perp }k_{\perp
}^{2}+D_{\parallel }k_{z}^{2}.  \label{2.24}
\end{equation}

\subsection{Nonequilibrium Impurities Structure Factor}

Let us now consider the effect produced by the concentration gradient $%
\nabla c$. The nonequilibrium part of\ $S\left( \overrightarrow{k},\omega
\right) $contains contributions arising from $J_{i}^{F}$, which is not
coupled with $\delta v_{i}$, $\delta n_{i}$, and three different
contributions arising from the dynamics of the nematic solvent which are
expressed as director, velocity and cross director-velocity correlation
functions, that is,
\begin{equation}
S^{neq}\left( \overrightarrow{k},\omega \right) =S_{J}^{neq}\left(
\overrightarrow{k},\omega \right) +S_{nn}^{neq}\left( \overrightarrow{k}%
,\omega \right) +S_{nv}^{neq}\left( \overrightarrow{k},\omega \right)
+S_{vv}^{neq}\left( \overrightarrow{k},\omega \right) .  \label{2.25}
\end{equation}%
$S_{J}^{neq}\left( \overrightarrow{k},\omega \right) $ is the nonequilibrium
contribution arising from $J_{i}^{F}$ due to the assumption of the validity
of the local version of the fluctuation dissipation theorem (\ref{2.9}),
which is given by Eq. (30) in $I$,
\begin{align}
S_{J}^{neq}\left( \overrightarrow{k},\omega \right) & \equiv \left\langle
\delta \widetilde{c}\left( \overrightarrow{k},\omega \right) \delta
\widetilde{c}\left( -\overrightarrow{k},-\omega \right) \right\rangle ^{neq}
\notag \\
& =-2\left\langle \delta \widetilde{c}\left( \overrightarrow{k},\omega
\right) \delta \widetilde{c}\left( -\overrightarrow{k},-\omega \right)
\right\rangle ^{eq}  \notag \\
& \times \omega \frac{\left\vert \nabla c\right\vert }{c^{o}}\left(
D_{\parallel }k_{z}\cos \psi +D_{\perp }k_{\perp }\sin \psi \right)
\left\vert \widetilde{G}\left( \overrightarrow{k},\omega \right) \right\vert
^{2}.  \label{4.13}
\end{align}%
Furthermore, since longitudinal and transverse fluctuations are uncoupled,
\begin{eqnarray}
S_{nn}^{neq}\left( \overrightarrow{k},\omega \right)
&=&-D_{a}^{2}\left\vert G\left( \overrightarrow{k},\omega \right)
\right\vert ^{2}\left\{ k_{z}^{2}a_{1}^{2}\left( \overrightarrow{k}\right)
\left\langle \delta \tilde{n}_{1}\left( \overrightarrow{k},\omega \right)
\delta \tilde{n}_{1}\left( -\overrightarrow{k},-\omega \right) \right\rangle
\right.   \notag \\
&&+\left[ 2k_{z}a_{3}\left( \overrightarrow{k}\right) +\left( k_{\perp }-%
\frac{k_{z}^{2}}{k_{\perp }}\right) a_{2}\left( \overrightarrow{k}\right) %
\right] ^{2}  \notag \\
&&\left. \times \left\langle \delta \tilde{n}_{3}\left( \overrightarrow{k}%
,\omega \right) \delta \tilde{n}_{3}\left( -\overrightarrow{k},-\omega
\right) \right\rangle \right\} ,  \label{2.26}
\end{eqnarray}%
\begin{eqnarray}
S_{nv}^{neq}\left( \overrightarrow{k},\omega \right)
&=&D_{a}\left\vert G\left( \overrightarrow{k},\omega \right)
\right\vert ^{2}\left\{ -2k_{z}a_{1}^{2}\left(
\overrightarrow{k}\right) \rm{Im}\left\{ \left\langle \delta
\tilde{n}_{1}\left( \overrightarrow{k},\omega \right) \delta
\tilde{v}_{1}\left( -\overrightarrow{k},-\omega \right)
\right\rangle
\right\} \right.   \notag \\
&&+2c_{o}k\left[ 2k_{z}a_{3}\left( \overrightarrow{k}\right) +\left(
1-\left( \frac{k_{z}}{k_{\perp }}\right) ^{2}\right) k_{\perp }a_{2}\left(
\overrightarrow{k}\right) \right]   \notag \\
&&\times \rm{Re}\left\{ \left\langle \delta \tilde{n}_{3}\left(
\overrightarrow{k},\omega \right) \delta \tilde{v}_{3}\left( -%
\overrightarrow{k},-\omega \right) \right\rangle \right\}   \notag \\
&&+2a_{2}\left( \overrightarrow{k}\right) \left[ 2k_{z}a_{3}\left(
\overrightarrow{k}\right) +\left( 1-\left( \frac{k_{z}}{k_{\perp }}\right)
^{2}\right) k_{\perp }a_{2}\left( \overrightarrow{k}\right) \right]   \notag
\\
&&\left. \times \Im \left\{ \left\langle \delta
\tilde{v}_{2}\left(
\overrightarrow{k},\omega \right) \delta \tilde{n}_{3}\left( -%
\overrightarrow{k},-\omega \right) \right\rangle \right\} \right\} ,
\label{2.27}
\end{eqnarray}%
\begin{eqnarray}
S_{vv}^{neq}\left( \overrightarrow{k},\omega \right)  &=&\left\vert G\left(
\overrightarrow{k},\omega \right) \right\vert ^{2}\left\{ -a_{1}^{2}\left(
\overrightarrow{k}\right) \left\langle \delta \tilde{v}_{1}\left(
\overrightarrow{k},\omega \right) \delta \tilde{v}_{1}\left( -%
\overrightarrow{k},-\omega \right) \right\rangle \right.   \notag \\
&&+a_{2}^{2}\left( \overrightarrow{k}\right) \left\langle \delta \tilde{v}%
_{2}\left( \overrightarrow{k},\omega \right) \delta \tilde{v}_{2}\left( -%
\overrightarrow{k},-\omega \right) \right\rangle   \notag \\
&&\left. -2c_{o}ka_{2}\left( \overrightarrow{k}\right)
\rm{Im}\left\{ \left\langle \delta \tilde{v}_{2}\left(
\overrightarrow{k},\omega \right) \delta \tilde{v}_{3}\left(
-\overrightarrow{k},-\omega \right) \right\rangle \right\}
\right\} ,  \label{2.28}
\end{eqnarray}%
where $a_{\mu }\left( \overrightarrow{k}\right) $, $\mu =1,2,3,$ represent
the transverse and longitudinal components of the concentration gradient,
defined in a similar fashion as $\delta \tilde{n}_{\mu }$ and $\delta \tilde{%
v}_{\mu }$ (Eqs.(\ref{2.18b})-(\ref{2.22})).\ It should be pointed out that
Eqs.(\ref{2.26})-(\ref{2.28}) show that in contrast to equilibrium, in $NESS$
the density gradient introduces a coupling between the concentration
fluctuations of the solute and the velocity and orientation equilibrium
fluctuations of the solvent. These contributions should be calculated by
first evaluating the required correlation functions of the solvent.

\section{Solvent equilibrium correlation functions}

Let us recall that the hydrodynamic fluctuations of a thermotropic nematic
evolve on three widely separated time-scales corresponding to the the
relaxation of orientational, visco-heat and sound modes, respectively, \cite%
{rodriguez2}, \cite{creta}. These relaxation times are such that $\tau
_{orientation}\sim \nu /Kk^{2}$, $\tau _{visco-heat}\sim \rho c_{p}/\kappa
k^{2}\sim \rho /\nu k^{2}$ and $\tau _{sound}\sim 1/c_{s}k$ , where $\nu $
denotes any of the nematic's viscosities, $K$ is the elastic constant, $c_{p}
$ denotes the specific heat at constant pressure, $\kappa $ is the magnitude
of any of the components of the thermal conductivity tensor and $c_{s}$ is
the isentropic sound speed of the nematic. For values of $k$ corresponding
to a hydrodynamic description and for typical values of the material
parameters of a thermotropic nematic \cite{khoo}, the following relation
holds%
\begin{equation}
\tau _{orientation}\gg \tau _{visco-heat}\gg \tau _{sound}.  \label{2.29}
\end{equation}%
By estimating the order of magnitude of the elements of the hydrodynamic
matrices of the time evolution equations for the nematic's fluctuations,
which are given in Ref.\ \cite{playa} by Eqs. (24), (25), (38)-(42), it is
possible\ to identify the following groups of variables $\left\{ \delta
\tilde{n}_{1},\delta \tilde{n}_{3}\right\} $, $\left\{ \delta \hat{v}%
_{1},\delta \hat{v}_{2},\delta \hat{T}\right\} $, $\left\{ \delta \hat{v}%
_{3},\delta \hat{p}\right\} $ as slow, semi-slow and fast, respectively. The
wide separation between these time-scales will now be exploited to eliminate
the faster variables from the general dynamical equations obtaining, thus, a
reduced description in which only the slower variables are involved. For
this purpose we use the time-scaling perturbation method developed in Refs.
\cite{geigen1}, \cite{geigen3}, which allows to find a contracted
description in terms of the slow variables only. The corresponding reduced
dynamical matrix will be constructed by a perturbation procedure, where the
perturbation parameters are the ratios $\tau _{visco-heat}/\tau
_{orientation}$ and $\tau _{sound}/\tau _{visco-heat},$\cite{hijar}. Using
this formalism it can be shown that in the slow time-scale, the director
fluctuations $\delta \tilde{n}_{1}$ and $\delta \tilde{n}_{3}$ obey the
stochastic equations%
\begin{equation}
-i\omega \delta \tilde{n}_{\mu }=-\omega _{n\mu }\left( \overrightarrow{k}%
\right) \delta \tilde{n}_{\mu }-\tilde{\sigma}_{n\mu },\text{ }\mu =1,3,
\label{2.30}
\end{equation}%
where the fluctuating terms $\tilde{\sigma}_{n\mu }$ are
\begin{equation}
\tilde{\sigma}_{n_{1}}\equiv \frac{1}{k_{\perp }}\left[ k_{x}\tilde{\Upsilon}%
_{y}-k_{y}\tilde{\Upsilon}_{x}+\frac{1}{2}\frac{\left( 1+\lambda \right)
k_{z}}{\nu _{2}k_{\perp }^{2}+\nu _{3}k_{z}^{2}}\left( k_{x}k_{j}\tilde{%
\Sigma}_{yj}-k_{y}k_{j}\tilde{\Sigma}_{yj}\right) \right] ,  \label{2.30a}
\end{equation}%
\begin{equation}
\tilde{\sigma}_{n_{3}}\equiv \frac{1}{k}\left[ k_{x}\tilde{\Upsilon}%
_{x}+k_{y}\tilde{\Upsilon}_{y}+\frac{1}{2}\frac{\left( 1+\lambda \right)
k_{z}^{2}+\left( 1-\lambda \right) k_{\perp }^{2}}{\nu _{3}k_{\perp
}^{4}+2\left( \nu _{1}+\nu _{2}-\nu _{3}\right) k_{\perp }^{2}k_{z}^{2}+\nu
_{3}k_{z}^{4}}\left( k^{2}k_{j}\tilde{\Sigma}_{zj}-k_{z}k_{i}k_{j}\tilde{%
\Sigma}_{ij}\right) \right] .  \label{2.30b}
\end{equation}%
Here $\Upsilon _{i}$ and $\Sigma _{ij}$ denote the stochastic components of
the director's quasi-current and the stress tensor, $\nu _{ijkl}$, which
obey the the fluctuation-dissipation relations \cite{rodriguez2}%
\begin{equation}
\left\langle \Sigma _{ij}\left( \overrightarrow{r},t\right) \Sigma
_{kl}\left( \overrightarrow{r}^{\prime },t^{\prime }\right) \right\rangle
=2k_{B}T_{o}\nu _{ijkl}\delta \left( \overrightarrow{r}^{\prime }-%
\overrightarrow{r}\right) \delta \left( t^{\prime }-t\right) ,  \label{2.30c}
\end{equation}%
\begin{equation}
\left\langle \Upsilon _{i}\left( \overrightarrow{r},t\right) \Upsilon
_{j}\left( \overrightarrow{r}^{\prime },t^{\prime }\right) \right\rangle =%
\frac{2k_{B}T_{o}}{\gamma _{1}}\delta _{ij}^{\perp }\delta \left(
\overrightarrow{r}^{\prime }-\overrightarrow{r}\right) \delta \left(
t^{\prime }-t\right) ,  \label{2.30d}
\end{equation}%
where $k_{B}$ is Boltzmann's constant, $\gamma _{1}$ is the orientational
viscosity coefficient. The viscosity tensor is
\begin{eqnarray}
\nu _{ijkl} &=&\nu _{2}(\delta _{jl}\delta _{ik+}\delta _{il}\delta
_{jk})+2(\nu _{1}+\nu _{2}-2\nu _{3})n_{i}^{o}n_{j}^{o}n_{k}^{o}n_{l}^{o}
\notag \\
&&+(\nu _{3}-\nu _{2})(n_{j}^{o}n_{l}^{o}\delta
_{ik}+n_{j}^{o}n_{k}^{o}\delta _{il}+n_{i}^{o}n_{k}^{o}\delta
_{jl}+n_{i}^{o}n_{l}^{o}\delta _{jk})  \notag \\
&&+(\nu _{4}-\nu _{2})\delta _{ij}\delta _{kl}+(\nu _{5}-\nu _{4}+\nu
_{2})(\delta _{ij}n_{k}^{o}n_{l}^{o}+\delta _{kl}n_{i}^{o}n_{j}^{o})\text{,}
\label{2.30e}
\end{eqnarray}%
where $\nu _{1}$, $\nu _{2}$ and $\nu _{3}$ are three shear viscosity
coefficients and $\nu _{5}$ and $\nu _{4}-\nu _{2}$ denote two bulk
viscosity coefficients. The quantity
\begin{equation}
\delta _{ij}^{\perp }\equiv \delta _{ij}-n_{i}^{o}n_{j}^{o}  \label{2.31}
\end{equation}%
is a projection operator and $\delta _{ij}$ denotes a Kronecker delta. In
the above equations we have also used the following abbreviations%
\begin{equation}
\omega _{n_{1}}\left( \overrightarrow{k}\right) =\frac{1}{\gamma _{1}}\left(
K_{2}k_{\perp }^{2}+K_{3}k_{z}^{2}\right) \left[ 1+\frac{1}{4}\frac{\gamma
_{1}\left( 1+\lambda \right) ^{2}k_{z}^{2}}{\nu _{2}k_{\perp }^{2}+\nu
_{3}k_{z}^{2}}\right] ,  \label{2.31a}
\end{equation}%
\begin{equation}
\omega _{n_{3}}\left( \overrightarrow{k}\right) =\frac{1}{\gamma _{1}}\left(
K_{1}k_{\perp }^{2}+K_{3}k_{z}^{2}\right) \left\{ 1+\frac{1}{4}\frac{\gamma
_{1}\left[ \left( 1+\lambda \right) k_{z}^{2}+\left( 1-\lambda \right)
k_{\perp }^{2}\right] ^{2}}{\nu _{3}k_{\perp }^{4}+2\left( \nu _{1}+\nu
_{2}-\nu _{3}\right) k_{\perp }^{2}k_{z}^{2}+\nu _{3}k_{z}^{4}}\right\} ,
\label{2.32}
\end{equation}%
where $K_{1}$, $K_{2}$ and $K_{3}$ are, respectively, the splay, twist and
bend elastic constants and $\lambda $ is a non-dissipative coefficient
associated with the director's relaxation.

From Eq. (31) in Ref. \cite{playa}, we obtain the reduced equation for the
semi-slow transverse variable, $\delta \tilde{v}_{1}$,%
\begin{equation}
-i\omega \delta \tilde{v}_{1}=-\omega _{v_{1}}\left( \overrightarrow{k}%
\right) \delta \tilde{v}_{1}-\tilde{\sigma}_{v_{1}},  \label{2.32a}
\end{equation}%
where the stochastic force term $\tilde{\sigma}_{v_{1}}$ is given by%
\begin{equation}
\tilde{\sigma}_{v_{1}}=\frac{i}{\rho _{o}k_{\perp }}\left( k_{x}k_{j}\tilde{%
\Sigma}_{yj}-k_{y}k_{j}\tilde{\Sigma}_{yj}\right)   \label{2.32b}
\end{equation}%
and the reduced characteristic frequency $\omega _{v_{1}}\left(
\overrightarrow{k}\right) $ is%
\begin{equation}
\omega _{v_{1}}\left( \overrightarrow{k}\right) =\frac{1}{\rho _{o}}\left(
\nu _{2}k_{\perp }^{2}+\nu _{3}k_{z}^{2}\right) .  \label{2.32c}
\end{equation}%
Similarly, from Eqs. (39)-(42) in Ref. \cite{playa}, we obtain the
corresponding equations for the semi-slow longitudinal variables $\delta
\tilde{v}_{2}$ and $\delta \tilde{T}$,
\begin{equation}
-i\omega \left(
\begin{array}{c}
\delta \tilde{v}_{2} \\
\delta \tilde{T}%
\end{array}%
\right) =-\left(
\begin{array}{cc}
\omega _{v2}\left( \overrightarrow{k}\right)  & 0 \\
0 & \omega _{T}\left( \overrightarrow{k}\right)
\end{array}%
\right) \left(
\begin{array}{c}
\delta \tilde{v}_{2} \\
\delta \tilde{T}%
\end{array}%
\right) -\left(
\begin{array}{c}
\tilde{\sigma}_{v2} \\
\tilde{\sigma}_{T}%
\end{array}%
\right) ,  \label{2.32d}
\end{equation}%
with%
\begin{equation}
\tilde{\sigma}_{v2}=\frac{i}{\rho _{o}k_{\perp }}\left[ \frac{k}{k_{\perp }}%
k_{i}\tilde{\Sigma}_{zi}-\frac{k_{z}}{k_{\perp }k}k_{i}k_{j}\tilde{\Sigma}%
_{ij}\right] ,  \label{2.32e}
\end{equation}%
\begin{equation}
\tilde{\sigma}_{T}=i\frac{1}{\rho _{o}c_{p}}k_{i}\tilde{Q}_{i}.
\label{2.32f}
\end{equation}%
Here $Q_{i}$ is the stochastic heat flux which satisfies the
fluctuation-dissipation theorem%
\begin{equation}
\left\langle Q_{i}\left( \overrightarrow{r},t\right) Q_{j}\left(
\overrightarrow{r}^{\prime },t^{\prime }\right) \right\rangle
=2k_{B}T_{o}^{2}\kappa _{ij}\delta \left( \overrightarrow{r}^{\prime }-%
\overrightarrow{r}\right) \delta \left( t^{\prime }-t\right) ,  \label{2.35a}
\end{equation}%
$\kappa _{ij}=\kappa _{\perp }\delta _{ij}+\kappa _{a}n_{i}^{o}n_{j}^{o}$
with $\kappa _{a}\equiv \kappa _{\parallel }-\kappa _{\perp }$, is the heat
conductivity tensor and $\omega _{v_{2}}\left( \overrightarrow{k}\right) $,
$\omega _{_{T}}\left( \overrightarrow{k}\right) $ are given by%
\begin{equation}
\omega _{v_{2}}\left( \overrightarrow{k}\right) \equiv \frac{1}{\rho
_{o}k^{2}}\left[ \nu _{3}\left( k_{z}^{4}+k_{\perp }^{4}\right) +2\left( \nu
_{1}+\nu _{2}-\nu _{3}\right) k_{z}^{2}k_{\perp }^{2}\right] ,  \label{2.35b}
\end{equation}%
\begin{equation}
\omega _{_{T}}\left( \overrightarrow{k}\right) \equiv D_{\perp }^{T}k_{\perp
}^{2}+D_{\parallel }^{T}k_{z}^{2},  \label{2.35c}
\end{equation}%
where $D_{\perp }^{T}\equiv \kappa _{\perp }/\rho _{o}c_{p}$, $D_{\parallel
}^{T}\equiv \kappa _{\parallel }/\rho _{o}c_{p}$, stand for the thermal
diffusivity coefficients of the nematic along the directions parallel and
perpendicular to $\hat{n}^{o}$.

Analogously, from Eqs. (60) in Ref. \cite{playa}, we obtain the reduced
equation for the fast variables $\delta \tilde{v}_{3}$ and $\delta \tilde{p}$
\begin{equation}
-i\omega \left(
\begin{array}{c}
\delta \tilde{v}_{3} \\
\delta \tilde{p}%
\end{array}%
\right) =-\left(
\begin{array}{cc}
\omega _{v3}\left( \overrightarrow{k}\right)  & ic_{s}k\left( i\frac{\chi
_{T}}{\gamma \rho }\right) ^{1/2} \\
ic_{s}k\left( \frac{\gamma \rho }{\chi _{T}}\right) ^{1/2} & \left( \gamma
-1\right) \omega _{T}\left( \overrightarrow{k}\right)
\end{array}%
\right) \left(
\begin{array}{c}
\delta \tilde{v}_{3} \\
\delta \tilde{p}%
\end{array}%
\right) -\left(
\begin{array}{c}
\tilde{\sigma}_{v3} \\
\tilde{\sigma}_{p}%
\end{array}%
\right) ,  \label{2.33}
\end{equation}%
which is valid in the fastest time scale. The stochastic noises $\tilde{%
\sigma}_{v_{3}}$ and $\tilde{\sigma}_{p}$ are defined by
\begin{equation}
\tilde{\sigma}_{v3}\equiv \frac{i}{\rho _{o}k}k_{i}k_{j}\tilde{\Sigma}_{ij},
\label{2.34}
\end{equation}%
\begin{equation}
\tilde{\sigma}_{p}\equiv ic_{s}\left[ \frac{\left( \gamma -1\right) }{%
c_{p}T_{o}}\right] ^{1/2}k_{i}\tilde{Q}_{i},  \label{2.35}
\end{equation}%
where we have used the definition
\begin{eqnarray}
\omega _{v_{3}}\left( \overrightarrow{k}\right)  &\equiv &\frac{1}{\rho
_{o}k^{2}}\left[ \left( \nu _{2}+\nu _{4}\right) k_{\perp }^{4}+2\left( 2\nu
_{3}+\nu _{5}\right) k_{\perp }^{2}k_{z}^{2}\right.   \notag \\
&&\left. +\left( 2\nu _{1}+\nu _{2}-\nu _{4}+2\nu _{5}\right) k_{z}^{4}
\right] .  \label{2.36}
\end{eqnarray}

It is essential to stress that the dynamic equation Eq. (\ref{2.30}) is
correct in the slowest time scale, that is, for times of the order of $\tau
_{orientation}$. Similarly, Eqs. (\ref{2.32a}) and (\ref{2.32d}) are valid
for times of the order of \ $\tau _{visco-heat}$\ and Eq. (\ref{2.33})
describes the dynamics of the fast variables in the fast time-scale
characterized for times of the order of $\tau _{sound}$. As a first
approximation we extrapolate Eqs.(\ref{2.32a}), (\ref{2.32d}) and (\ref{2.33}%
) to the slow time-scale, in order to calculate the required nematic's
correlation functions. Solving Eqs. (\ref{2.30}), (\ref{2.32a}), (\ref{2.32d}%
) and (\ref{2.33}) for $\delta \tilde{n}_{\mu }$ and $\delta \tilde{v}_{\mu }
$, using the definitions of the stochastic terms $\tilde{\sigma}_{n\mu }$, $%
\tilde{\sigma}_{v\mu }$, $\tilde{\sigma}_{T}$, $\tilde{\sigma}_{P}$ \ and
using the fluctuation dissipation relations (\ref{2.30c}), (\ref{2.30d}), (%
\ref{2.35a}), we arrive at%
\begin{equation}
\left\langle \delta \tilde{n}_{1}\left( \overrightarrow{k},\omega \right)
\delta \tilde{n}_{1}\left( -\overrightarrow{k},-\omega \right) \right\rangle
=-\frac{\epsilon }{\gamma _{1}}\frac{\alpha _{1}\left( \overrightarrow{k}%
\right) }{\omega ^{2}+\omega _{n_{1}}^{2}\left( \overrightarrow{k}\right) },
\label{sol30}
\end{equation}%
\begin{equation}
\left\langle \delta \tilde{n}_{3}\left( \overrightarrow{k},\omega \right)
\delta \tilde{n}_{3}\left( -\overrightarrow{k},-\omega \right) \right\rangle
=-\frac{\epsilon }{\gamma _{1}}\frac{k_{\perp }^{2}}{k^{2}}\frac{\alpha
_{3}\left( \overrightarrow{k}\right) }{\omega ^{2}+\omega _{n_{3}}^{2}\left(
\overrightarrow{k}\right) },  \label{sol31}
\end{equation}%
\begin{equation}
\left\langle \delta \tilde{v}_{1}\left( \overrightarrow{k},\omega \right)
\delta \tilde{v}_{1}\left( -\overrightarrow{k},-\omega \right) \right\rangle
=-\frac{\epsilon }{\rho _{o}}\frac{\omega _{v_{1}}\left( \overrightarrow{k}%
\right) }{\omega ^{2}+\omega _{v_{1}}^{2}\left( \overrightarrow{k}\right) },
\label{4.6f}
\end{equation}%
\begin{equation}
\left\langle \delta \tilde{v}_{2}\left( \overrightarrow{k},\omega \right)
\delta \tilde{v}_{2}\left( -\overrightarrow{k},-\omega \right) \right\rangle
=\frac{\epsilon }{\rho _{o}}\frac{\omega _{v_{2}}\left( \overrightarrow{k}%
\right) }{\omega ^{2}+\omega _{v_{2}}^{2}\left( \overrightarrow{k}\right) }
\label{4.6g}
\end{equation}%
and%
\begin{eqnarray}
\left\langle \delta \hat{v}_{3}\left( \overrightarrow{k},\omega \right)
\delta \hat{v}_{3}\left( -\overrightarrow{k},-\omega \right) \right\rangle
&=&-\frac{\epsilon }{\rho _{o}}\frac{1}{\left[ \left( \omega +c_{s}k\right)
^{2}+\Gamma ^{2}\left( \overrightarrow{k}\right) \right] \left[ \left(
\omega -c_{s}k\right) ^{2}+\Gamma ^{2}\left( \overrightarrow{k}\right) %
\right] }  \notag \\
&&\times \left\{ \omega _{v}\left( \overrightarrow{k}\right) \left[ \omega
^{2}+\left( \gamma -1\right) ^{2}\omega _{T}^{2}\left( \overrightarrow{k}%
\right) \right] \right.   \notag \\
&&\left. +c_{s}^{2}k^{2}\left( \gamma -1\right) \omega _{_{T}}\left(
\overrightarrow{k}\right) \right\} ,  \label{4.6z}
\end{eqnarray}%
where $\epsilon \equiv 2\left( 2\pi \right) ^{4}\delta ^{4}\left( 0\right)
k_{B}T_{o}$,
\begin{equation}
\alpha _{1}(\overrightarrow{k})=1+\frac{\gamma _{1}\left( 1+\lambda
^{2}\right) k_{z}^{2}}{4\left( \nu _{2}k_{\perp }^{2}+\nu
_{3}k_{z}^{2}\right) },  \label{4.6d}
\end{equation}%
\begin{equation}
\alpha _{3}(\overrightarrow{k})=1+\frac{\gamma _{1}\left[ (1+\lambda
)k_{z}^{2}+(1-\lambda )k_{\perp }^{2}\right] ^{2}}{4\left[ \nu _{3}\left(
k_{\perp }^{4}+k_{z}^{4}\right) +2\left( \nu _{1}+\nu _{2}-\nu _{3}\right)
k_{\perp }^{2}k_{z}^{2}\right] },  \label{4.6e}
\end{equation}%
\begin{equation}
\Gamma \left( \overrightarrow{k}\right) =\frac{1}{2}\left[ \omega
_{v_{3}}\left( \overrightarrow{k}\right) +\left( \gamma -1\right) \omega
_{_{T}}\left( \overrightarrow{k}\right) \right] .  \label{2.43}
\end{equation}%
$\Gamma \left( \overrightarrow{k}\right) $is the anisotropic sound
attenuation coefficient of the nematic. To arrive at the previous
correlation functions we took into account that $\Upsilon _{i}$, $\Sigma
_{ij}$ and $Q_{i}$ are not correlated, and that for typical values of \ the
material parameters of a thermotropic nematic the relation $c_{s}k\gg \omega
_{v\mu }\left( \overrightarrow{k}\right) ,\omega _{T}\left( \overrightarrow{k%
}\right) \gg \omega _{n\mu }\left( \overrightarrow{k}\right) $, which are
equivalent to (\ref{2.29}), holds.

Following the same procedure described above, it can be shown that%
\begin{eqnarray}
\rm{Im}\left\{ \left\langle \delta \tilde{n}_{1}\left( \overrightarrow{k}%
,\omega \right) \delta \tilde{v}_{1}\left( -\overrightarrow{k},-\omega
\right) \right\rangle \right\}  &=&-\frac{\epsilon }{\rho _{o}}k_{z}\left(
1+\lambda \right) \frac{\omega ^{2}}{\omega _{v1}^{2}\left( \overrightarrow{k%
}\right) }  \notag \\
&&\times \left[ \frac{1}{\omega ^{2}+\omega _{n1}^{2}\left( \overrightarrow{k%
}\right) }-\frac{1}{\omega ^{2}+\omega _{v1}^{2}\left( \overrightarrow{k}%
\right) }\right] ,  \label{2.43a}
\end{eqnarray}%
\begin{eqnarray}
\rm{Re}\left\{ \left\langle \delta \tilde{n}_{3}\left( \overrightarrow{k}%
,\omega \right) \delta \tilde{v}_{3}\left( -\overrightarrow{k},-\omega
\right) \right\rangle \right\}  &=&-\frac{\epsilon }{2\rho _{o}}\frac{%
k_{\perp }}{k^{2}}\left[ \left( 1+\lambda \right) k_{z}^{2}+\left( 1-\lambda
\right) k_{\perp }^{2}\right]   \notag \\
&&\times \left\{ \frac{\left( \gamma -1\right) \omega _{T}\left(
\overrightarrow{k}\right) }{\omega ^{2}+\omega _{n3}^{2}\left(
\overrightarrow{k}\right) }+\frac{1}{2}\left[ \frac{\Gamma \left(
\overrightarrow{k}\right) }{\left( \omega +c_{s}k\right) ^{2}+\Gamma
^{2}\left( \overrightarrow{k}\right) }\right. \right.   \notag \\
&&\left. \left. -\frac{\Gamma \left( \overrightarrow{k}\right) }{\left(
\omega -c_{s}k\right) ^{2}+\Gamma ^{2}\left( \overrightarrow{k}\right) }%
\right] \right\} ,  \label{2.43c}
\end{eqnarray}%
\begin{eqnarray}
\rm{Im}\left\{ \left\langle \delta \tilde{v}_{2}\left( \overrightarrow{k}%
,\omega \right) \delta \tilde{n}_{3}\left( -\overrightarrow{k},-\omega
\right) \right\rangle \right\}  &=&-\frac{\epsilon }{2\rho _{o}}\frac{\left(
1+\lambda \right) k_{z}^{2}+\left( 1-\lambda \right) k_{\perp }^{2}}{k^{2}}%
\frac{k_{\perp }\omega ^{2}}{\omega _{v2}^{2}\left( \overrightarrow{k}%
\right) }  \notag \\
&&\times \left[ \frac{1}{\omega ^{2}+\omega _{n3}^{2}\left( \overrightarrow{k%
}\right) }-\frac{1}{\omega ^{2}+\omega _{v_{2}}^{2}\left( \overrightarrow{k}%
\right) }\right] ,  \label{2.43d}
\end{eqnarray}%
\begin{eqnarray}
\rm{Im}\left\{ \left\langle \delta \tilde{v}_{2}\left( \overrightarrow{k}%
,\omega \right) \delta \tilde{v}_{3}\left( -\overrightarrow{k},-\omega
\right) \right\rangle \right\}  &=&-\frac{\epsilon }{\rho _{o}}\frac{%
k_{z}k_{\perp }}{k^{2}}\frac{\omega \omega _{a}\left( \overrightarrow{k}%
\right) }{c_{s}^{2}k^{2}}\left\{ \frac{\left( \gamma -1\right) \omega
_{T}\left( \overrightarrow{k}\right) +\omega _{v2}\left( \overrightarrow{k}%
\right) }{\omega ^{2}+\omega _{v_{2}}^{2}\left( \overrightarrow{k}\right) }%
\right.   \notag \\
&&+\frac{1}{2}\left[ \frac{\Gamma \left( \overrightarrow{k}\right) }{\left(
\omega +c_{s}k\right) ^{2}+\Gamma ^{2}\left( \overrightarrow{k}\right) }%
\right.   \notag \\
&&\left. \left. -\frac{\Gamma \left( \overrightarrow{k}\right) }{\left(
\omega -c_{s}k\right) ^{2}+\Gamma ^{2}\left( \overrightarrow{k}\right) }%
\right] \right\} .  \label{2.43e}
\end{eqnarray}%
These expressions determine the required nematic's correlation functions,
Eqs. (\ref{2.26})-(\ref{2.28}).

\section{Results}

In \emph{I} we showed that in equilibrium, the main contribution to the
dynamic structure factor of the impurities is a central Rayleigh lorentzian
peak. However, when compressibility effects are considered, the equilibrium
dynamic structure factor of the impurities involves the equilibrium
auto-correlation function of the fluctuating component of the velocity along
$\overrightarrow{k}$. In previous work we have shown that this correlation
function contains information about the propagating sound modes of the
nematic which gives raise to its Brillouin peaks \cite{rodriguez}.
Therefore, it can be expected that the light scattering spectrum of the
impurities will also show these features and it should also exhibit two
Brillouin-like peaks. For this reason, hereafter we will only consider the
behavior of the dynamic structure factor, $S\left( \overrightarrow{k},\omega
\right) $, for frequencies close to $\omega =0$ and $\omega =c_{s}k$.

The evaluation of the different contributions of $S\left( \overrightarrow{k}%
,\omega \right) $ can be simplified by considering the order of magnitude of
the involved material parameters. The diffusion coefficients of dyes in a
thermotropic nematic are of the order of $D\sim 10^{-7}$ $cm^{2}$ $s^{-1}$
\cite{rondelez}, while for a typical room temperature thermotropic we have $%
\rho _{o}\sim 1$ $g$ $cm^{-3}$, $c_{s}\sim 10^{5}$ $cm$ $s^{-1}$, $\nu \sim
10^{-1}$ $poise$, $D^{T}\sim 10^{-3}$ $cm^{2}$ $s^{-1}$ and $K\sim 10^{-7}$ $%
dyn$. At low concentrations ($5\%$) $c_{o}\sim 10^{20}$ $cm^{-3}$. This
implies that the characteristic diffusion time of the impurities is much
slower than the corresponding one to the director relaxation and, therefore,
to all the other dynamic processes, $\omega _{n\mu }\left( \overrightarrow{k}%
\right) \gg \omega _{D}\left( \overrightarrow{k}\right) $. Therefore, by
inserting Eqs. (\ref{sol30})-(\ref{2.43e}) into Eqs. (\ref{2.26})-(\ref{2.28}%
) and retaining only the leading terms corresponding with the previous
orders of magnitude of the material parameters at $\omega \simeq \omega
_{D}\left( \overrightarrow{k}\right) $ (Rayleigh peak, $R$) and at $\omega
\simeq c_{s}k$ (Brillouin peaks $B$), we obtain explicit expressions for the
different contributions of $S\left( \overrightarrow{k},\omega \right) $
which are given below.

\subsection{Equilibrium Light Scattering Spectrum}

\subsubsection{Central Peak}

In order to compare the relative effect of the compressible character of the
solvent and the external concentration gradient on the spectrum of the
impurities, we define the dimensionless structure factor $\bar{S}\left(
\overrightarrow{k},\omega \right) $ by
\begin{equation}
\bar{S}\left( \overrightarrow{k},\omega \right) \equiv \frac{S\left(
\overrightarrow{k},\omega \right) }{S_{in}^{eq}\left( \overrightarrow{k}%
,0\right) },  \label{2.44}
\end{equation}%
where $S_{in}^{eq}\left( \overrightarrow{k},\omega \right) $ represents the
structure factor of an incompressible nematic in the equilibrium state. When
the incompressibility condition is implemented in Eq. (\ref{2.23}) we obtain
\begin{equation}
S_{in}^{eq}\left( \overrightarrow{k},\omega \right) \equiv 2\left( 2\pi
\right) ^{4}\delta ^{4}\left( 0\right) c_{o}\frac{\omega _{D}\left(
\overrightarrow{k}\right) }{\omega ^{2}+\omega _{D}^{2}\left(
\overrightarrow{k}\right) }.  \label{2.44a}
\end{equation}

From Eqs. (\ref{2.23}) and (\ref{4.6z}) it follows that the equilibrium
dynamic structure factor of the impurities for small frequency shifts, i.e. $%
\omega \simeq \omega _{D}\left( \overrightarrow{k}\right) $, is a Lorentzian
given by%
\begin{equation}
\bar{S}_{R}^{eq}\left( \omega _{o}\right) =\frac{1}{1+\omega _{o}^{2}}\left[
1+\frac{c_{o}k_{B}T_{o}}{\rho _{o}c_{s}^{2}}\frac{\left( \gamma -1\right)
\omega _{T}\left( \overrightarrow{k}\right) }{\omega _{D}\left(
\overrightarrow{k}\right) }\right] ,  \label{2.45}
\end{equation}%
where the normalized frequency $\omega _{o}\equiv \omega /\omega _{D}\left(
\overrightarrow{k}\right) $ has been used. The second term on the r. h. s.
represents the contribution due to the nematic's compressibility. This can
be seen more clearly if this term is rewritten in terms of the isentropic
compressibility defined by the thermodynamic relation $\chi _{s}=1/\rho
_{o}c_{s}^{2}$. The value of this term can be estimated by taking typical
values of the involved parameters. Indeed, in the case of the diffusion of
two different dyes (methylred and nitrozo di-methyl aniline) at the room
temperature in the thermotropic nematic $MBBA$ at low concentrations ($5\%$%
), we have%
\begin{equation}
\frac{c_{o}k_{B}T_{o}}{\rho _{o}c_{s}^{2}}\frac{\left( \gamma -1\right)
\omega _{T}\left( \overrightarrow{k}\right) }{\omega _{D}\left(
\overrightarrow{k}\right) }\sim 10^{-1},  \label{2.46}
\end{equation}%
which implies that the compressibility contribution to the central peak may
be significant and of the order of $\sim 10\%$. To illustrate this effect
quantitatively we consider a fixed $k$ corresponding to an incident wave
with $k_{1}=10^{5}$ $cm^{-1}$ for a scattering angle $\theta =1%
%TCIMACRO{\U{b0}}%
%BeginExpansion
{{}^\circ}%
%EndExpansion
$, in the scattering geometry of Fig.1. If we plot $\bar{S}_{R}^{eq}$ for
both, the incompressible and compressible cases as functions of $\omega _{o}$%
, we get the curves shown in Fig. 2. Note that the dynamic structure factor
in the compressible case (continuos line) is higher and wider than in the
incompressible situation (dashed line). For $c_{o}=2\times 10^{19}$ $cm^{-3}$
(diluted suspension at $1\%$) the relative differences of the height and
half width at half height are $5\%$ and $2.5\%$, respectively. For $%
c_{o}=1\times 10^{20}$ $cm^{-3}$ (diluted suspension at $5\%$) this changes
are $25\%$ and $11\%$ , which may be significant.

\begin{figure}[tbp]
\includegraphics[width=11cm]{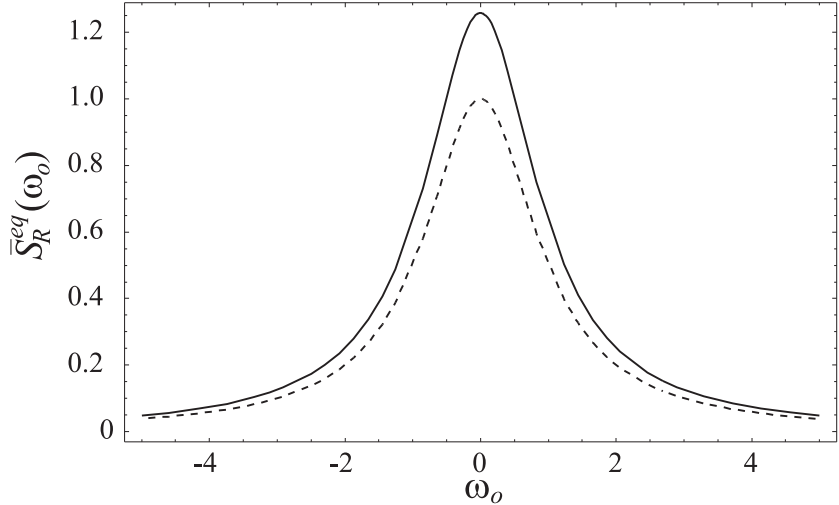}
\caption{Normalized central peak in equilibrium, $\bar{S}_{R}^{eq}$, as
function of the normalized frequency $\protect\omega _{o}$. (\textbf{---})
corresponds to a compressible nematic solvent and (- - -) denotes the
incompressible contribution obtained in $I$. $\bar{S}_{R}^{eq}$ is
calculated from Eq. (\protect\ref{2.45}) for typical values of the material
parameters and the scattering geometry shown in Fig. 1 with $k_{1}=1\times
10^{5}$ $cm^{-1}$ and $\protect\theta =1^{\circ }$.}
\end{figure}

\subsubsection{Brillouin Peaks}

The Brillouin peaks are located at the frequencies $\omega \simeq \pm $ $%
c_{s}k$ and Eqs. (\ref{2.23}) and (\ref{4.6z}) yield the following
expression for the normalized Brillouin spectrum of the impurities in terms
of $\omega _{o}$
\begin{equation}
\bar{S}_{B}^{eq}\left( \omega _{o}\right) =\frac{1}{2}\frac{k_{B}T_{o}c_{o}}{%
\rho _{o}c^{2}}\Gamma _{o}\left[ \frac{1}{\left( \omega _{o}+\frac{c_{s}k}{%
\omega _{D}}\right) ^{2}+\Gamma _{o}^{2}}+\frac{1}{\left( \omega _{o}-\frac{%
c_{s}k}{\omega _{D}}\right) ^{2}+\Gamma _{o}^{2}}\right] ,  \label{2.47}
\end{equation}%
with $\Gamma _{o}\equiv \Gamma (\overrightarrow{k})/\omega _{D}\left(
\overrightarrow{k}\right) $. First, notice that the ratio of the maxima of
the central and Brillouin peaks is
\begin{equation}
\zeta =\frac{k_{B}T_{o}c_{o}}{\rho _{o}c_{s}^{2}\Gamma _{o}}.  \label{2.48}
\end{equation}%
Thus, if we consider the order of magnitude of the involved parameters for
the diffusion of dyes in a typical thermotropic nematic as above, it follows
that $\zeta \sim 10^{-10}$. This shows that the Brilloiun component of the
light scattering spectrum of the impurities is negligible when compared in
front of its central part.

\subsection{Nonequilibrium Light Scattering Spectrum}

We now consider the effect of the concentration gradient on the dynamic
structure factor of the impurities when the solvent is in its nematic phase.
It is convenient to introduce the normalized concentration gradient
components by%
\begin{equation}
\bar{a}_{i}\equiv \frac{a_{i}}{k_{1}c_{o}}.  \label{3.1}
\end{equation}

\subsubsection{Central Peak}

If we keep only the dominant terms Eqs. (\ref{2.26})-(\ref{2.28}), for $\bar{%
a}_{i}$ in the range $10^{-8}<\bar{a}_{i}<1$, we obtain that at low
frequencies, $\omega \simeq \omega _{D}\left( \overrightarrow{k}\right) $, $%
S_{vv}^{neq}\left( \overrightarrow{k},\omega \right) \gg S_{nn}^{neq}\left(
\overrightarrow{k},\omega \right) \gg S_{nv}^{neq}\left( \overrightarrow{k}%
,\omega \right) $ and the leading nonequilibrium contribution to the dynamic
structure factor can then be written in the form%
\begin{equation}
\bar{S}_{R}^{neq}\left( \omega _{o}\right) =\frac{c_{o}k_{B}T_{o}}{\rho
_{o}\omega _{D}\left( \overrightarrow{k}\right) }k_{1}^{2}\frac{1}{1+\omega
_{o}^{2}}\left[ \frac{\bar{a}_{1}^{2}\left( \overrightarrow{k}\right) }{%
\omega _{v1}\left( \overrightarrow{k}\right) }+\frac{\bar{a}_{2}^{2}\left(
\overrightarrow{k}\right) }{\omega _{v2}\left( \overrightarrow{k}\right) }%
\right] .  \label{3.2}
\end{equation}%
Thus, in the nonequilibrium state the central peak of the dynamic structure
factor, $\bar{S}_{R}=\bar{S}_{R}^{eq}+\bar{S}_{R}^{neq}$, takes the form%
\begin{eqnarray}
\bar{S}_{R}\left( \omega _{o}\right)  &=&\frac{1}{1+\omega _{o}^{2}}\left\{
1+\frac{c_{o}k_{B}T_{o}}{\rho _{o}c_{s}^{2}}\frac{\left( \gamma -1\right)
\omega _{T}\left( \overrightarrow{k}\right) }{\omega _{D}\left(
\overrightarrow{k}\right) }\right.   \notag \\
&&\left. +\frac{c_{o}k_{B}T_{o}}{\rho _{o}\omega _{D}\left( \overrightarrow{k%
}\right) }k_{1}^{2}\left[ \frac{\bar{a}_{1}^{2}\left( \overrightarrow{k}%
\right) }{\omega _{v_{1}}\left( \overrightarrow{k}\right) }+\frac{\bar{a}%
_{2}^{2}\left( \overrightarrow{k}\right) }{\omega _{v_{2}}\left(
\overrightarrow{k}\right) }\right] \right\} .  \label{3.3}
\end{eqnarray}%
This result indicates that the effect of the concentration gradient
increases both, the height and the half-width at half-height of the
spectrum. The relative magnitude of the nonequilibrium contribution is
measured by the function
\begin{equation}
\xi =\frac{\bar{S}_{R}-\bar{S}_{R}^{eq}}{\bar{S}_{R}^{eq}}=\frac{%
c_{o}k_{B}T_{o}}{\rho _{o}\omega _{D}}k_{1}^{2}\left[ \frac{\bar{a}_{1}^{2}}{%
\omega _{v_{1}}}+\frac{\bar{a}_{2}^{2}}{\omega _{v_{2}}}\right]   \label{3.4}
\end{equation}%
If we take $c_{o}\sim 10^{20}$ $cm^{-3}$, $T_{o}\sim 300$ $K$, $D\sim 10^{-7}
$ $cm^{2}$ $s^{-1}$, $\nu \sim 10^{-1}$ $poise$, $\rho _{o}\sim 1$ $gcm^{-3}$%
, $k_{1}\sim 10^{5}$ $cm^{-1}$ and small scattering angles, $\theta \sim 1%
%TCIMACRO{\U{b0}}%
%BeginExpansion
{{}^\circ}%
%EndExpansion
$, a typical value is
\begin{equation}
\xi \sim \frac{c_{o}k_{B}T_{o}}{\rho _{o}D\nu k^{4}}k_{1}^{2}\bar{a}^{2}\sim
10^{12}\bar{a}^{2}.  \label{3.5}
\end{equation}%
This suggests that the nonequilibrium contribution could be significant, $%
\sim $ $10\%$, for normalized concentration gradients as small as $\bar{a}%
\sim 10^{-6}$, which are four orders of magnitude smaller than those
gradients used in \emph{I}. In Fig. 3 we compare $\bar{S}_{R}\left( \omega
_{o}\right) $ with its equilibrium component $\bar{S}_{R}^{eq}\left( \omega
_{o}\right) $ for the diffusion of dyes in the thermotropic nematic $MBBA$
at low concentrations ($5\%$), for the following specific values of the
normalized concentration gradient, $a_{x}=a_{y}=0$, $a_{z}=1\times 10^{-6}$,
and for the scattering process shown in Fig. 1 with $k_{1}=10^{5}$ $cm^{-1}$%
, $\theta =1%
%TCIMACRO{\U{b0}}%
%BeginExpansion
{{}^\circ}%
%EndExpansion
$. We notice that the spectrum becomes higher and wider when the
concentration gradient is present (continuos line) than in the equilibrium
case (dashed line). For the considered values of the involved quantities the
increment in the height is about $25\%$ and the change in the half-width at
half-height is $\sim 12\%$.

\begin{figure}[tbp]
\includegraphics[width=11cm]{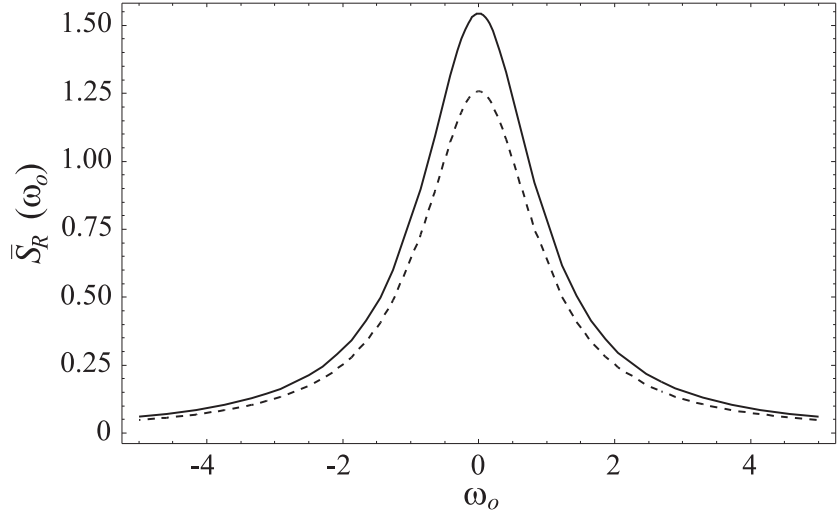}
\caption{Normalized central peak $\bar{S}_{R}\left( \protect\omega %
_{o}\right) $ as given by Eq. (\protect\ref{3.3}) for $k_{1}=1\times 10^{5}$
$cm^{-1}$ and $\protect\theta =1^{\circ }$. (- - -) represents the
equilibrium contribution. (---) denotes the dynamic structure factor in the $%
NESS$ induced by a normalized concentration gradient $\protect%
\overrightarrow{a}\equiv (a_{x},a_{y},a_{z})$ with $a_{x}=a_{y}=0$, $%
a_{z}=1\times 10^{-6}$ ($\protect\psi =0^{\circ }$).}
\end{figure}

\subsubsection{Brillouin Peaks}

Following the same procedure, we find that the leading contribution to the
nonequilibrium part of the dynamic structure factor at $\omega \simeq \pm
c_{s}k$ arises from $S_{nv}^{neq}\left( \overrightarrow{k},\omega \right) $.
More specifically, from the cross correlation $\left\langle \delta \tilde{n}%
_{3}\left( \overrightarrow{k},\omega \right) \delta \tilde{v}_{3}\left( -%
\overrightarrow{k},-\omega \right) \right\rangle $ and the normalized form
given by (\ref{2.44}) we get
\begin{eqnarray}
\bar{S}_{B}^{neq}\left( \omega _{o}\right)  &=&-\frac{c_{o}k_{B}T_{o}D_{a}}{%
\rho _{o}c_{s}^{2}k^{3}}k_{1}k_{\perp }^{2}\left[ 2k_{z}\overline{a}%
_{3}\left( \overrightarrow{k}\right) +\left( 1-\left( \frac{k_{z}}{k_{\perp }%
}\right) ^{2}\right) k_{\perp }\overline{a}_{2}\left( \overrightarrow{k}%
\right) \right]   \notag \\
&&\times \frac{k_{z}A\left( \overrightarrow{k}\right) \left( \gamma
-1\right) \omega _{T}\left( \overrightarrow{k}\right) }{\omega
_{n3}^{2}\left( \overrightarrow{k}\right) }\frac{\omega _{o}}{1+\omega
_{o}^{2}},  \label{3.6}
\end{eqnarray}%
where%
\begin{eqnarray}
A\left( \overrightarrow{k}\right)  &\equiv &\frac{(1+\lambda
)k_{z}^{2}+(1-\lambda )k_{\perp }^{2}}{\nu _{3}\left( k_{\perp
}^{4}+k_{z}^{4}\right) +2\left( \nu _{1}+\nu _{2}-\nu _{3}\right) k_{\perp
}^{2}k_{z}^{2}}\left[ \left( 2\nu _{3}+\nu _{5}-\nu _{2}-\nu _{4}\right)
k_{\perp }^{2}\right.   \notag \\
&&\left. +\left( 2\nu _{1}+\nu _{2}-2\nu _{3}-\nu _{4}+\nu _{5}\right)
k_{z}^{2}\right] .  \label{4.6j}
\end{eqnarray}%
Notice that the nonequilibrium term is an odd function of the frequency and,
as a consequence, the external concentration gradient induces an asymmetry
in the spectrum in such a way that one of its Brillouin peaks increases
while the other decreases in the same amount with respect to their
equilibrium counterparts. This effect is linear in the concentration
gradient magnitude. Furthermore, since it can be readily shown that the
function $\bar{S}_{B}^{neq}$ does not varies considerably over the frequency
intervals of the order of $\Gamma \left( \overrightarrow{k}\right) $ around $%
\pm c_{s}k$, we can make the approximation
\begin{eqnarray}
\bar{S}_{B}^{neq}\left( \omega _{o}\right)  &=&\mp \frac{c_{o}k_{B}T_{o}D_{a}%
}{\rho _{o}c_{s}^{3}k^{4}}k_{1}k_{\perp }^{2}\left[ 2k_{z}\overline{a}%
_{3}+\left( 1-\left( \frac{k_{z}}{k_{\perp }}\right) ^{2}\right) k_{\perp }%
\overline{a}_{2}\right]   \notag \\
&&\times \frac{k_{z}A\left( \gamma -1\right) \omega _{T}}{\omega _{n3}^{2}},
\label{4.6k}
\end{eqnarray}%
where the upper sign corresponds to the peak located at $-c_{s}k$ and the
lower sing to the peak at $c_{s}k$. The relative magnitude of this effect is
given by the quantity
\begin{equation}
\chi \equiv \frac{\bar{S}_{B}-\bar{S}_{B}^{eq}}{\bar{S}_{B}^{eq}}=A\frac{%
k_{z}k_{1}k_{\perp }^{2}}{k^{3}}\left[ 2k_{z}\overline{a}_{3}+\left(
1-\left( \frac{k_{z}}{k_{\perp }}\right) ^{2}\right) k_{\perp }\overline{a}%
_{2}\right] \frac{\omega _{o}\overline{\omega }_{T}}{\omega _{n_{3}}^{2}},
\label{4.6l}
\end{equation}%
whose significance can be estimated by taking into account the order of
magnitude of the involved parameters and introducing the normalized gradient
components according to (\ref{3.1}). In this way we find that%
\begin{equation}
\chi \sim 10^{-10}k_{1}\bar{a}\text{ }cm.  \label{4.6m}
\end{equation}%
This implies that the nonequilibrium contribution to the Brillouin part of
the spectrum could be significant only for normalized gradients of the order
of $\bar{a}\sim 10^{-2}$, which are much larger than those considered for
the central peak. Moreover, if the angular dependence of $\chi $\ is taken
into account, it turns out that this contribution is actually one order of
magnitude smaller.

In order to complete our analysis we now compare the normalized Brillouin
component of the dynamic structure factor of the impurities, $\bar{S}%
_{B}\left( \overrightarrow{k},\omega \right) $, with respect to $\bar{S}%
_{B}^{eq}\left( \overrightarrow{k},\omega \right) $, using the same values
of the material parameters as before. For instance, for a normalized
concentration gradient with components $a_{x}=a_{y}=1\times 10^{-1}$, $%
a_{z}=0$, and the scattering process shown in Fig. 1 with $k_{1}=10^{5}$ $%
cm^{-1}$ and $\theta =90%
%TCIMACRO{\U{b0}}%
%BeginExpansion
{{}^\circ}%
%EndExpansion
$, the height of the Brillouin peak located at $\omega =-c_{s}k$ increases $%
\sim $ $7\%$ while the one located at $\omega =c_{s}k$ decreases by the same
amount, as shown in Fig. 4. Finally, we stress that both, the Brillouin
component of the spectrum and the nonequilibrium effect on the Brillouin
peaks are several orders of magnitude smaller than the central component and
the nonequilibrium effect on this peak, respectively. Thus, the possible
experimental observation of the effects discussed in this work seems to be
more feasible for the central component of $S\left( \overrightarrow{k}%
,\omega \right) $.

\begin{figure}[tbp]
\includegraphics{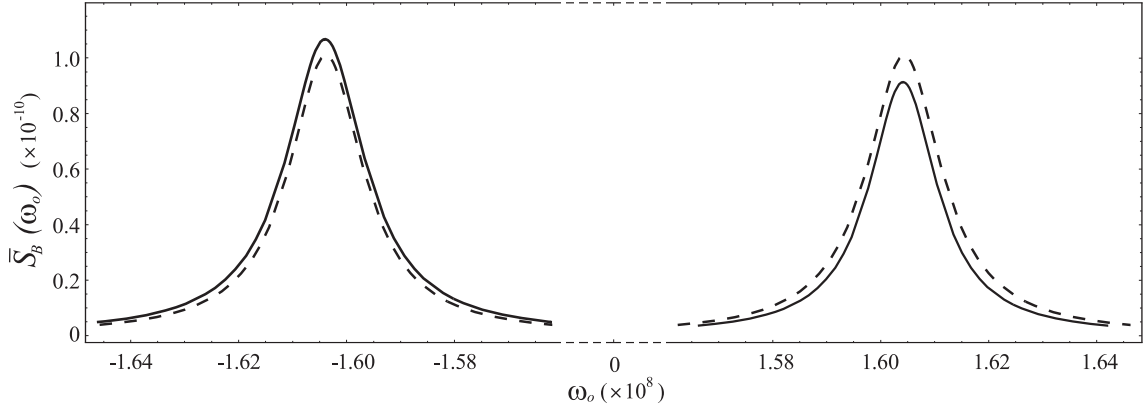}
\caption{Normalized Brillouin spectrum $\bar{S}_{R}\left( \protect\omega %
_{o}\right) $ as given by Eqs. (\protect\ref{2.47}) and (\protect\ref{3.6})
for $k_{1}=1\times 10^{5}$ $cm^{-1}$ and $\protect\theta =90^{\circ }$. (- -
-) represents the equilibrium part of the spectrum while (---) is the
dynamic structure factor in the $NESS$ induced by a normalized concentartion
gradient with components $a_{x}=a_{y}=1\times 10^{-1}$, $a_{z}=0$.}
\end{figure}

\section{Concluding Remarks}

Summarizing, by using a fluctuating hydrodynamic approach we have
investigated theoretically the influence of the effects produced by a
uniform impurities concentration gradient on the light scattering spectrum
of a suspension in a compressible nematic solvent. We compared both cases,
when the solvent is in a fully thermodynamic equilibrium state and in a
non-equilibrium steady state induced by a dye-concentration gradient. In the
former state, the spectrum is symmetric (Lorentzian) with respect to the
frequency shifts, but anisotropic through its explicit dependence on the
diffusion coefficients of the dye, parallel and normal to the mean molecular
axis of the nematic. The values of these coefficients were taken from
experimental measurements of diffusion of methylred and nitrozo di-methyl
aniline in a $MBBA$ solvent. Our results showed that the compressibility
increases the height and the width at mid-height with respect to the
incompressible case in amounts which vary up to $25\%$ for a dye diluted
suspension at $5\%$ in $MBBA$.

As was discussed above, the nonequilibrium correction turns out to be
several orders of manitude larger for the central peak of the spectrum than
for the Brillouin part. The Rayleigh peak becomes higher and wider when the
concentration gradient is present with respect to the equilibrium case. For
the considered values of the involved quantities, the increment in the
height is about $25\%$ and the change in the half-width at half-height is $%
12\%$, as indicated in Fig. 3. The size of this effect depends on the square
of the gradient components.

To our knowledge, the physical situation dealt with here has not been
considered in the literature and our model calculations yield new results
that might be observable; however, this remains to be assessed.

\begin{acknowledgments}
Partial financial support from DGAPA-UNAM IN108006 and from FENOMEC through
grant CONACYT 400316-5-G25427E, Mexico, is gratefully acknowledged.
\end{acknowledgments}


\begin{thebibliography}{99}
\bibitem{suspension1} H. H\'{\i}jar and R. F. Rodr\'{\i}guez, \textit{Phys.
Rev. E}, \textbf{69}, 051701 (2004)

\bibitem{beysens} D. Beysens, T. Garrabos and G. Zalczer, \textit{Phys. Rev.
Lett.\ }\textbf{48},\textbf{\ }403 (1980)

\bibitem{sengers} J. R. Dorfman, T. R. Kirkpatrick and J. V. Sengers,
\textit{Annu. Rev. Phys. Chem.} \textbf{45}, 213 (1994)

\bibitem{law1} B. Law, M., P. N. Segr\`{e}, R. W. Gammon and J. V. Sengers,
\textit{Phys. Rev. A} \textbf{41},\textbf{\ }816 (1990)

\bibitem{rondelez} F. Rondelez, \textit{Sol. State Commun}. \textbf{14}, 815%
\textbf{\ }(1974)

\bibitem{janig} F. J\"{a}hnig and H. Scmidt, \textit{Ann. Phys\ (New York) }%
\textbf{71}, 129 (1972)

\bibitem{franklin} W. Franklin, \textit{Phys. Rev. A\ }\textbf{11}, 2156
(1975)

\bibitem{landau} L. D. Landau and E. Lifshitz, \textit{Fluid Dynamics}
(Pergamon, New York, 1959)

\bibitem{forster} D. Forster, \textit{Hydrodynamic Fluctuations, Broken
Symmetry and Correlation Functions} (Benjamin, Reading, 1975)

\bibitem{forster2} D. Forster, T. Lubensky, P. C. Martin, J. Swift, P. S.
Pershan, Phys. Rev. Lett. \textbf{26}, 1016 (1971)

\bibitem{creta} R. F. Rodr\'{\i}guez and H. H\'{\i}jar, \textit{Eur. Phys.
J. B} \textbf{50}, 105-110 (2006)

\bibitem{khoo} I. C. Khoo and S. T. Wu, \textit{Optics and Nonlinear Optics
of Liquid Crystals} (World Scientific, Singapore, 1993)

\bibitem{geigen1} U. Geigenm\"{u}ller, U. M. Titulaer and B. U. Felderhof,
Physica, \textbf{119A}, 41 (1983)

\bibitem{geigen3} U. Geigenm\"{u}ller, B. U. Felderhof and U. M. Titulaer,
Physica, \textbf{120A}, 635 (1983)

\bibitem{hijar} H. H\'{\i}jar, Ph. D. Dissertation, National University of
Mexico (2006) (in Spanish)

\bibitem{rodriguez2} J. F. Camacho, H. H\'{\i}jar and R. F. Rodr\'{\i}guez,
\textit{Physica A} \textbf{348}, 252 (2005)

\bibitem{playa} H. H\'{\i}jar and R. F. Rodr\'{\i}guez, \textit{Rev. Mex.
Fis.}, (2006) (in press)

\bibitem{rodriguez} R. F. Rodr\'{\i}guez and J. F. Camacho, Nonequilibrium
thermal light scattering from nematic liquid crystals in \textit{Recent
Developments in Mathematical and Experimental Physics, Vol. B Statistical
Physics and Beyond}, A. Macias, E. D\'{\i}az and F. Uribe, editors (Kluwer,
New York, 2002) pp. 209-224
\end{thebibliography}
\end{document}